\documentclass[conference,10pt]{IEEEtran}
\IEEEoverridecommandlockouts
\usepackage{cite}
\usepackage{amsmath,amssymb,amsfonts}
\usepackage{algorithmic}
\usepackage{graphicx}
\usepackage{textcomp}
\usepackage{nicefrac,url}
\usepackage{tikz, pgfplots}
\usetikzlibrary{positioning,calc}
\usepackage{xcolor,booktabs,multirow}

\def\BibTeX{{\rm B\kern-.05em{\sc i\kern-.025em b}\kern-.08em
    T\kern-.1667em\lower.7ex\hbox{E}\kern-.125emX}}

\newcommand{\PO}{\mathcal{P}_0}
\newcommand{\PI}{\mathcal{P}_1}

\newcommand{\Natural}{natural}

\newcommand{\secret}{secret}

\newcommand{\Naturals}{\Natural~}

\newcommand{\secrets}{\secret~}

\DeclareMathOperator*{\argmin}{arg\,min}
\definecolor{darkgreen}{RGB}{0,128,0}

\begin{document}

\title{Detecting Adversarial Examples -\\ A Lesson from Multimedia Forensics
\thanks{This research was funded by Archimedes Privatstiftung, Innsbruck, Austria and Deutsche Forschungsgemeinschaft (DFG) under grant ``Informationstheoretische Schranken digitaler Bildforensik''.}
}

\author{
\IEEEauthorblockN{Pascal Sch\"{o}ttle, Alexander Schl\"{o}gl, Cecilia Pasquini, and Rainer B\"{o}hme}
\IEEEauthorblockA{\textit{Department of Computer Science, University of Innsbruck}, Innsbruck, Austria \\
\{pascal.schoettle; alexander.schloegl; cecilia.pasquini; rainer.boehme\}@uibk.ac.at}
}

\maketitle

\begin{abstract}
Adversarial classification is the task of performing robust classification in the presence of a strategic attacker. 
Originating from information hiding and multimedia forensics, adversarial classification recently received a lot of attention in a broader security context. 
In the domain of machine learning-based image classification, adversarial classification can be interpreted as detecting so-called adversarial examples, which are slightly altered versions of benign images. They are specifically crafted to be misclassified with a very high probability by the classifier under attack. 
Neural networks, which dominate among modern image classifiers, have been shown to be especially vulnerable to these adversarial examples.

However, detecting subtle changes in digital images has always been the goal of multimedia forensics and steganalysis. In this paper, we highlight the parallels between these two fields and secure machine learning. 

Furthermore, we adapt a linear filter, similar to early steganalysis methods, to detect adversarial examples that are generated with the projected gradient descent (PGD) method, the state-of-the-art algorithm for this task.
We test our method on the MNIST database and show for several parameter combinations of PGD that our method can reliably detect adversarial examples. 

Additionally, the combination of adversarial re-training and our detection method effectively reduces the attack surface of attacks against neural networks.
Thus, we conclude that adversarial examples for image classification possibly do not withstand detection methods from steganalysis, and future work should explore the effectiveness of known techniques from multimedia forensics in other adversarial settings.
\end{abstract}

\begin{IEEEkeywords}
Adversarial Classification, Adversarial Examples, Multimedia Forensics, Steganalysis
\end{IEEEkeywords}

\section{Introduction}

The task of \emph{adversarial classification} is to perform robust and reliable classification in the presence of strategic attackers~\cite{Dritsoula17}. The nature of a strategic attacker is that she will not 
disregard knowledge about possible defense mechanisms. Rather, she adapts her attack strategy to circumvent the most probable defense mechanisms~\cite{Barni13}.

In machine learning-based classification, state-of-the-art attacks are so-called \emph{adversarial examples}~\cite{Szegedy13}. Adversarial examples are benign inputs that have been strategically modified by an attacker such that they are misclassified with a very high probability and confidence. Initially, adversarial examples were generated against classifiers based on \emph{convolutional neural networks} (CNNs), but soon it was shown that they generalize to other machine learning algorithms as well~\cite{Papernot16c}. 

Adversarial classification against adversarial examples can be achieved in two different ways: either the designers of the CNNs try to detect adversarial examples as adversarial (\emph{adversarial detection}) 
or they try to increase the robustness of the CNN in such a way that adversarial examples are classified in the class of the underlying benign example (\emph{robust classification}). 
But, to this day, no method to detect adversarial examples effectively exists and earlier work from the area of \emph{adversarial machine learning}~\cite{Barreno2006, Biggio2012} did not prove to be useful against adversarial examples, either.

Although adversarial examples do not only exist for image classifiers (e.\,g., they also exist for malware classifiers~\cite{Grosse17}), the main body of work is performed for the area of digital images. 
Thus, we restrict ourselves to this domain. 

Every method for generating adversarial examples from benign images calculates which pixels should be modified by how much (restricted by a distortion constraint) to maximize the probability of a misclassification, e.\,g.,~\cite{Szegedy13, Papernot16b, Madry17, Carlini17}.

Detecting subtle malicious changes in digital images has always been the goal of multimedia forensics and steganalysis\footnote{Note that in steganography/steganalysis jargon usually the steganalyst is the attacker and the steganographer is the defender. We refrain from statements about who is good or bad, but reverse their roles in this paper.}.  Without explicitly using the term adversarial classification, both domains perform adversarial detection  since the very beginning of scientific research in either of the fields. The strategic nature of the attackers here is defined by research in counter-forensics and steganography~\cite{Kirchner08}. Both are implicitly aware of possible detection methods and try to evade them.

In the field of digital image forensics~\cite{Sencar13}, a forensic investigator has to decide if a given image was manipulated by an image forger or not. Oftentimes, the image forger manipulates large connected parts of the image with the aim to change its semantic~\cite{Kirchner08}. Thereby, she might use \emph{tamper hiding} techniques~\cite{Kirchner07} to conceal traces she expects the forensic investigator to identify.

In steganalysis, a steganalyst has to decide if a given image has a message embedded by a steganographer~\cite{Fridrich09}. While embedding her message, the steganographer tries to modify individual pixel values in such a way that the steganalyst gets the least information about the fact that a message is embedded. To achieve this, every modern steganographic algorithm defines an adaptivity criterion that identifies which pixels are most suitable for embedding.

Machine learning-based approaches, and especially detectors built on CNNs, are by now very common in image forensics and steganalysis. But, to the best of our knowledge, nobody has tried to go the other way around, i.\,e., to use established methods from multimedia forensics or steganalysis to detect adversarial examples against CNNs. 

We close this gap and give an intuition on why, how, and what the area of secure machine learning can learn from the fields of multimedia forensics and steganalysis. As mentioned above, adversarial examples are generated by changing individual pixels of a benign image. As this is more similar to the embedding process in steganography than to the forgery creation that image forensics has to detect, we restrict our explanations to the field of steganalysis in the rest of the paper and develop a steganalysis-inspired linear prediction method to detect adversarial examples that are generated with the \emph{projected gradient descent} (PGD) method~\cite{Madry17}.

The remainder of the paper is organized as follows: Section~\ref{sec:back} gives the background about secure machine learning and steganalysis, and highlights the parallels and differences of these fields. 
As a proof of concept, we develop our method to detect adversarial examples in Section~\ref{sec:ws} and show its effectiveness in Section~\ref{sec:results}. 
Section~\ref{sec:conclusion} concludes.

\section{Background \& Parallels}
\label{sec:back}

The publication closest to our work is~\cite{Quiring17}. The authors show the parallels of attacks against and defenses for secure machine learning and digital watermarking, another subfield of multimedia security. We argue that the detection of adversarial examples rather falls into the domain of steganalysis and encourage researchers to make use of established steganalysis methods before they start out to reinvent the wheel.

\subsection{Secure Machine Learning}
\label{susec:advex}

The underlying assumption of every machine learning-based classifier is that the training data follows the same, possibly unknown distribution as the test data. For example, a supervised CNN-based classifier that has to distinguish $n$ different classes  is trained  with many samples $x_i$ and their corresponding labels $i \in \{1, 2, \ldots, n\}$. So, the CNN learns an approximation of the classification function $F(x_i) = i$, given a specific loss function $\ell_F(x_i,i)$, and predicts a label $i \in \{1, 2, \ldots, n\}$ for every sample encountered testing.

\subsubsection{Creating Adversarial Examples}
Intuitively, every (untargeted) adversarial examples starts from a benign example $x_i$. The attacker tries to find $r$, subject to a distortion constraint, such that $x_i+r$ gets misclassified by $F(\cdot)$. This can be achieved by solving the following  optimization problem:
\begin{align}\label{eq:advex}
\argmin_r &  ~d(x_i,x_i+r), \\ 
	\text{s. t. } & F(x_i+r) = i' \neq i, \nonumber
\end{align}
for a given distance metric $d$. 

For a given $x_i$, we define the $i$-th class as the \emph{benign class}, whose samples follow a distribution $\PI$. Accordingly, we denote as $\PO$ the distribution of samples belonging to every other class except $i$. This allows to transform every multi-class problem to the binary case With a slight abuse of notation, the goal of an attacker is to modify an image $x_1 \thicksim \PI$ such that it gets classified as drawn from $\PO$.

Among all the methods proposed for the generation of adversarial examples, e.\,g.,~\cite{Szegedy13,Papernot16b, Carlini17}, 
the so-called \emph{projected gradient descent} (PGD) method~\cite{Madry17} constitutes the state-of-the-art at the time of writing.

The PGD method basically is an iterated variant of the \emph{Fast Gradient Method} (FGM)~\cite{Goodfellow14}, which takes a single step of value $\alpha$ in the direction of the gradient of the loss function $\nabla \ell_F$. 
Additionally, all pixel values are clipped to the range of 0 and 1 to ensure a valid image in the end. PGD introduces a second variable $\varepsilon$ and sets $x^{[0]} = x_1$ to iteratively calculate
\begin{equation}
x^{[k+1]} = \text{clip}_{[x_1-\varepsilon,x_1+\varepsilon]}\big( \text{FGM}(x^{[k]})\big),
\label{eq:pgd}
\end{equation}
for a given number of iterations $K$. So, $x^{[K]}$ serves as an approximation of the optimal adversarial example $x_i+r$ in Eq.~\eqref{eq:advex}. The outer clipping ensures that $||x_1-x^{[K]}||_\infty \leq \varepsilon$ to model an attacker that is restricted by the infinity norm. Note that depending on the value of the gradient of the loss function the PGD method changes  individual pixel values up to a maximum of $\varepsilon$.

\subsubsection{Proposed Countermeasures}

Recent research on the countermeasures against adversarial examples mainly concentrates on increasing the robustness of the underlying neural network so that it classifies adversarial examples to the class of the benign object used for creating the example. One recent approach is to cut off the lower bit layers and only classifying the remaining image~\cite{Xu18}, in the hope that the adversarial modifications are concentrated in the lower bit layers. Another one is to re-train the neural network with adversarial examples~\cite{Madry17}, so-called \emph{adversarial re-training}.

Although the nature of adversarial examples is still not fully known~\cite{Tramer17}, it is accepted that they generalize over different networks: adversarial examples generated against one network are also likely to be misclassified by other networks~\cite{Moosavi17}.

\subsection{Steganalysis} 
In steganalysis, the distribution $\PO$ defines the distribution of all possible benign images (cover images), while $\PI$ is the distribution of stego images. Every modern steganographic embedding function defines a so-called adaptivity criterion which measures the distortion when changing single pixels in a specific way. During the embedding of the message, the steganographer tries to minimize the overall distortion with the goal that the stego image $x_1\thicksim \PI$ gets classified by the steganalyst as drawn from $\PO$. 
The steganographer decides on a per-pixel basis about the changes she introduces. The goal of the steganalyst is to decide for a given image $x_i$, from which distribution it was drawn. As this is a classic task for a neural network classifier, it comes as no surprise that CNN-based steganalysis attracted a lot of attention recently.

\subsection{Parallels and Differences}
\label{susec:parallels}

We summarize the main parallels and differences of secure machine learning and steganalysis in Table~\ref{tab:par}.

The main difference is the nature of the distribution $\PI$. 
In secure machine learning, both distributions $\PO$ and $\PI$ are exogenously given from real-world examples, and the goal of the attacker is to modify images drawn from $\PI$ so that they are classified as being drawn from $\PO$. This can be regarded as moving $x_1$ across the decision boundary as far as possible under the given distortion constraint, e.\,g., $\varepsilon$ for PGD.

Contrary to that, in steganalysis the distribution $\PI$ is given by the images created by the steganographer, and can thus be influenced by her attack algorithm. So, the goal of a steganographer is to create a distribution $\PI$ similar enough to $\PO$, such that a steganalyst cannot reliably differentiate between objects drawn from $\PI$ and objects drawn from $\PO$.

A higher value of $\varepsilon$ in PGD enables an attacker to move her adversarial examples farther into the space of $\PO$ by changing individual pixels more. By doing so, the attacker makes it harder even for adversarially re-trained networks to correctly classify adversarial examples~\cite{Madry17}.

But, detecting objects that deviate from an expected distribution is exactly what established methods from steganalysis are designed to do. So, the higher $\varepsilon$ for PGD is, and the less robust the CNN classifiers get, the better the performance of methods adapted from steganalysis should be.

\begin{table}
\caption{Parallels and differences of secure machine learning and steganalysis}
\begin{center}
\begin{tabular}{p{2.15cm}p{3cm}p{2.6cm}}
					& Secure Machine Learning					& Steganalysis						\\
\toprule
Attack point			& adversarial example						& stego image				\\
Decision by			& network (designer)							& steganalyst	 				\\

 Attack algorithm		& modification								& embedding						\\
 Attack parameters		& (internal) parameters						& adaptivity criterion		\\
 Attack surface			& individual pixels							& individual pixels		\\
 $\PO$: distribution of	& other class(es)							& cover images				\\
 $\PI$: distribution of		& benign class								& stego images				\\
 \multirow{2}{*}{Attacker's goal}				& get $x_1 \thicksim \PI$ classified 	& get $x_1 \thicksim \PI$ classified\\
 											&  as drawn from $\PO$				&  as drawn from $\PO$\\[1ex]
\midrule
 Nature of $\PI$			& exogenously given							& influenced by attacker	\\
\bottomrule
\end{tabular}
\label{tab:par}
\end{center}
\end{table}

\section{Proof-of-Concept: Adapting Steganalysis}
\label{sec:ws}

\subsection{Method}

One of the simplest and earliest steganalysis methods is based on the intuition that pixels that were changed during the embedding behave different than pixels that were not changed~\cite{Fridrich09}. For example, in a (unmodified) cover image, the original pixel values should be estimable from values of the surrounding pixels. If pixel $i$ was changed during embedding, its observed value $x^i$ will deviate from the estimated value $\hat{x}^i$. 
This estimation can be achieved by a simple linear filter of the following form~\cite{Ker08}:
\begin{align}
	 \hat{x}^i= x^i * \left( \begin{array}{ccc}
-\nicefrac{1}{4} & \nicefrac{1}{2} & -\nicefrac{1}{4} \\
\phantom{-}\nicefrac{1}{2} & 0 &\phantom{-} \nicefrac{1}{2} \\
-\nicefrac{1}{4} & \nicefrac{1}{2} & -\nicefrac{1}{4} \end{array} \right). \nonumber
\end{align}

Taking the average of the weighted differences over all $n$ pixels in an observed image can serve as an indicator if a message was embedded or not.
\begin{align}
	\hat{p} 
                        &= \frac{1}{n}\sum\limits_{i=0}^{n-1} w_i  \left( x^i - \hat{x}^i \right) \label{eq:adapuwws}
\end{align}

If $\hat{p}$ is relatively small, it can be expected that the image is a cover image. The weights $w_i$ in Eq.~\eqref{eq:adapuwws} account for local predictability, and one successful initialization~\cite{Ker08} suggest that $w_i^{-1} \propto 5+\sigma_i^2$ give accurate estimates, where $\sigma_i^2$ denotes the local variance in the neighborhood of pixel $i$ (but excluding the center pixel). It was shown that such an estimator, adapted to a specific way of changing the pixel values during embedding~\cite{Ker08}, coincides with an asymptotically uniformly most powerful (AUMP) hypothesis test~\cite{Fillatre12}.

\begin{figure}[b!]
    \begin{center}
        \begin{tikzpicture}[font=\small]
            \input{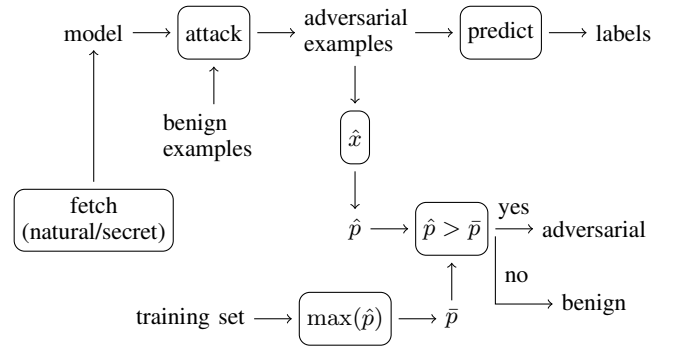}
        \end{tikzpicture}
        \caption{Block diagram of our experiments' workflow}
        \label{fig:blockdgr}
    \end{center}
\end{figure}

\subsection{Experimental Setup}
\label{susec:setup}

We test our method on the MNIST dataset\cite{Mnist} against adversarial examples generated by the PGD method~\cite{Madry17}. The MNIST dataset contains 60\,000 grayscale images of handwritten digits, which are split up into 50\,000 images in the \emph{training set} and 10\,000 images in the \emph{test set}. 

The detection of adversarial examples is performed by calculating $\hat{p}$ as in Eq.~\eqref{eq:adapuwws} for every image and classifying it as adversarial if $\hat{p}$ is above a certain threshold $\bar{p}$. We chose a conservative approach and set $\bar{p}$ to the maximum value obtained from the 50\,000 images from the training set, so that $\bar{p}$ ensures no false positives on the training set.

To ensure reproducibility, we did not train the CNNs ourselves but fetch the \emph{\Natural} and \emph{\secret} models from~\cite{Madry17}\footnote{Available at: https://github.com/MadryLab/mnist\_challenge}. Here, the \secrets model was re-trained with adversarial examples generated by PGD against the \Naturals model with $\varepsilon=0.3$.

Furthermore, we use the provided \emph{attack} script to generate adversarial examples for every image in the test set, for every $\varepsilon \in \{0.01, 0.02, \ldots, 0.5\}$ for both models.
This brings the total number of adversarial images used for testing to 100\,000.
All adversarial examples are generated with full knowledge of the CNN they are targeting, making this a white-box attack, and thus the worst-case scenario for a defender~\cite{Papernot18}. The overall setup of our experiments is depicted in Fig.~\ref{fig:blockdgr}.

\begin{table*}[ht!]
\caption{True positive rate of our detector (over all 10\,000 images of the MNIST test set)}
\begin{center}
\begin{tabular}{clllllllllllllll}
\toprule
\multirow{2}{*}{Model} & \multicolumn{15}{c}{$\varepsilon$}\\
	&	
0.010	&	0.050	&	0.100	&	0.150	&	0.200	&	0.250	&	0.300	&	0.330	&	0.350	&	0.370	&	0.400	&	0.420	&	0.450	&	0.470	&	0.500\\ \cmidrule{2-16}
natural	&	
0.000	&	0.000	&	0.529	&	0.982	&	1.000	&	1.000	&	1.000	&	1.000	&	1.000	&	1.000	&	1.000	&	1.000	&	1.000	&	1.000	&	1.000\\
secret	&	
0.000	&	0.000	&	0.000	&	0.001	&	0.009	&	0.025	&	0.047	&	0.137	&	0.320	&	0.574	&	0.848	&	0.937	&	0.985	&	0.995	&	0.999\\

\bottomrule
\end{tabular}
\label{tab:TPR}
\end{center}
\end{table*}

\begin{figure}[t!]
\begin{center}
\begin{tikzpicture}[>=stealth]
\begin{axis}[  	legend pos=south east,
       			axis x line=bottom,
			axis y line*=left,
        			xlabel=$\varepsilon$,
        			ylabel=True Positive Rate,  
        			ymax=1.005,
			ymin=-0.005,
        			xmin= 0,
			xmax=0.5,
			y label style={at={(0.05,0.5)}},
			]
\addplot[smooth,mark=none, darkgreen,] table[x index=0, y index=1] 
      {tpr_columns.dat};
\addlegendentry{\Natural}
\addplot[smooth,mark=none, red, ] table[x index=0, y index=3] 
       {tpr_columns.dat};
\addlegendentry{\secret}
\addplot +[mark=none,dotted, black] coordinates {(0.099, 0) (0.099, 1.005)};
\addplot +[mark=none,dashed, black] coordinates {(0.365, 0) (0.365, 1.005)};
\end{axis}
\end{tikzpicture}
\end{center}%
\caption{True positive rate of our method for different $\varepsilon$}
\label{fig:tpr}
\end{figure}
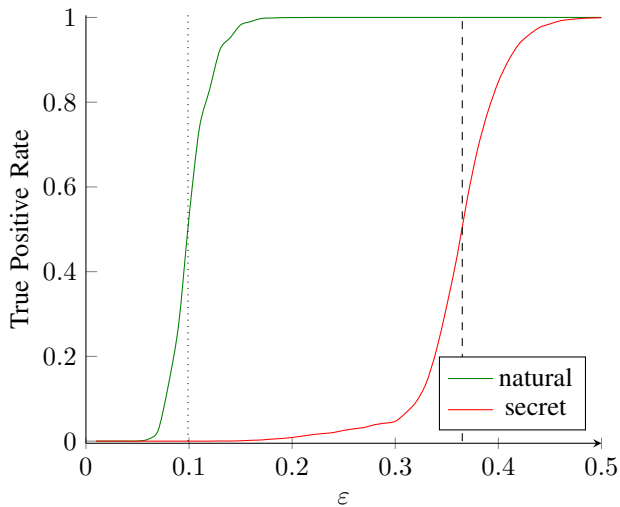

\section{Results}
\label{sec:results}

\subsection{Detecting Adversarial Examples}
As explained in Sec.~\ref{susec:setup}, our method is constructed in such a way that false positives are extremely unlikely and indeed, in none of our tests we encountered a single benign image that was classified as adversarial by our method. 

Thus, to fully assess the performance of our method, it is enough to report the true positive rate (TPR), i.\,e., the amount of adversarial examples that were correctly identified as adversarial by our method. 
Table~\ref{tab:TPR} lists the TPR for the whole test set for different values of $\varepsilon$. 
As we can see, our method improves for higher values of $\varepsilon$ for both models. As argued in Sec.~\ref{susec:parallels}, this is to be expected, as with increasing value of $\varepsilon$, the adversarial examples will deviate farther from $\PI$ and thus are better detectable by our method.  

It is interesting to observe that attacks against the adversarially re-trained model (\secret) are harder to detect for our method. First tests about the difference of the adversarial examples created against the re-trained model (omitted here due to space constraints) show that these adversarial examples change more pixel values in a homogeneous way, thus lying closer to our pixel prediction.

\begin{figure}[t!]
\begin{center}
\begin{tikzpicture}[>=stealth]
\begin{axis}[  	legend pos=north east,
       			axis x line=bottom,
			axis y line*=left,
        			xlabel=$\varepsilon$,
        			ylabel=Accuracy,  
        			ymax=1.005,
			ymin=-0.005,
        			xmin= 0,
			xmax=0.5,
			y label style={at={(0.05,0.5)}},
			]
\addplot[smooth,mark=none, darkgreen] table[x index=0, y index=1,] 
      {accuracies.dat};
\addlegendentry{\Natural}
\addplot[blue, smooth,mark=none, red, ] table[x index=0, y index=4] 
      {accuracies.dat};
\addlegendentry{\secret}
\addplot +[mark=none,dotted, black] coordinates {(0.099, 0) (0.099, 1.005)};
\addplot +[mark=none,dashed, black] coordinates {(0.365, 0) (0.365, 1.005)};
\end{axis} 
\end{tikzpicture}
\end{center}%
\caption{Accuracy of both classifiers from\cite{Madry17} for different $\varepsilon$}
\label{fig:acc}
\end{figure}
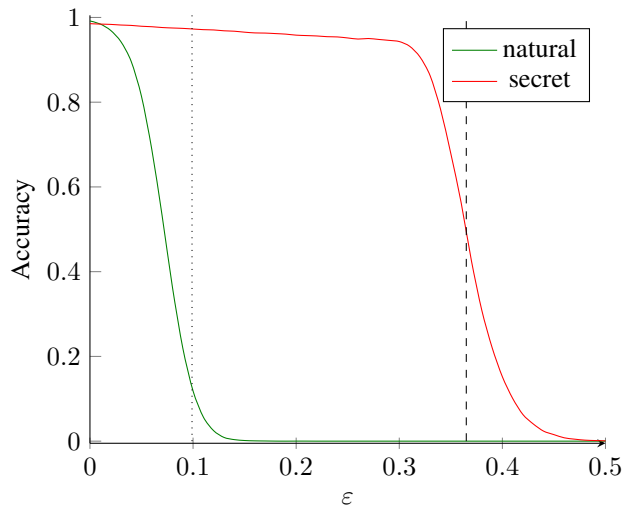

We plot the TPR for both models and all tested values of $\varepsilon$ in Fig.~\ref{fig:tpr}. In comparison to the accuracy of the models from~\cite{Madry17} (cf. Fig.~\ref{fig:acc}), it is observable, that our method indeed improves approximately at the point where the CNN classifiers lose robustness (the dotted/dashed vertical line indicates where our method achieves  $\approx 50\%$ TPR for the \Naturals (\secret) model). This orthogonal behavior of our adversarial detection and the robust classification obtained by adversarial re-training motivates a combination of both approaches.

\subsection{Combination of our method with adversarial re-training}
 
 Motivated by the observation from the previous section, we combined our method to detect adversarial examples with the adversarially re-trained CNNs from~\cite{Madry17}.
  
\begin{figure}[t!]
\begin{center}
\begin{tikzpicture}[>=stealth]
\begin{axis}[  	legend pos=south east,
       			axis x line=bottom,
			axis y line=left,
        			xlabel=$\varepsilon$,
        			ylabel=Accuracy,  
        			ymax=1.005,
			ymin=-0.005,
        			xmin= 0,
			xmax=0.5,
			y label style={at={(0.05,0.5)}},
			]
\addplot[smooth,mark=none, darkgreen] table[x index=0, y expr={\thisrowno{1}/10000}] 
      {eval.dat};
\addlegendentry{\Natural}
\addplot[smooth,mark=none, red] table[x index=0, y expr={\thisrowno{3}/10000}] 
      {eval.dat};
\addlegendentry{\secret}
\addplot +[mark=none,dotted, black] coordinates {(0.17, 0) (0.17, 1.005)};
\addplot +[mark=none,dashed, black] coordinates {(0.31, 0) (0.31, 1.005)};
\end{axis}
\end{tikzpicture}
\end{center}
\caption{Accuracy of the combination of both methods for different $\varepsilon$}
\label{fig:either}
\end{figure}
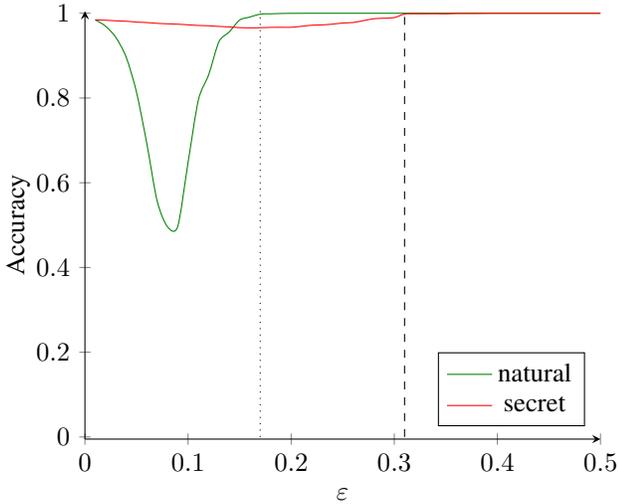

 For every image from the test set, we first create adversarial examples for every value of $\varepsilon$. Then, we apply our method to decide if the example is adversarial or not. All images that are not detected as adversarial by our method are handed to the CNN classifiers which predict their labels, see Fig.~\ref{fig:blockdgr}.  We plot the accuracies of the combined approach for both models in Fig.~\ref{fig:either}. As we can see, even for the natural model the overall accuracy never falls below 50\% and for the secret model the overall accuracy in never less then 96\%. This confirms that our method detects a majority of the adversarial examples that would have been misclassified by the CNNs. 
 
It is notable that for the natural model we achieve perfect separation of adversarial and benign samples for $\varepsilon \geq 0.17$ (dotted vertical line in Fig.~\ref{fig:either}), while this holds true for the secret model only for $\varepsilon \geq 0.31$ (dashed vertical line in Fig.~\ref{fig:either}).

\section{Conclusion}
\label{sec:conclusion}

Adversarial classification in the area of secure machine learning can roughly be divided into adversarial detection and robust classification. While the latter approach gained more attention recently, we argue in this paper that the detection of adversarial examples crafted against CNN-based classifiers can draw on long established methods from steganalysis. 

We highlight the conceptual parallels between the creation of adversarial examples and the generation of stego images and develop a very simple method to reliably detect adversarial examples generated by the PGD method. Furthermore, we give theoretical insights on why methods adapted from steganalysis can successfully complement robust classification: they are designed to perform well against exactly the adversarial examples that are hard to classify robustly.

An even better performance can be achieved by combining our method with adversarial re-trained CNNs. The minimum accuracy of the combined approach over all parameters is 96\%, almost at par with the accuracy of the tested CNNs for benign images. An additional benefit of the combined approach is that it efficiently reduces the freedom of an attacker, as it is very hard to defeat our method and adversarially re-trained CNNs with the same adversarial examples.

For our proof-of-concept, we restrict ourselves to a very simple method from the domain of steganalysis. Future work should identify, which methods from the field of multimedia forensics can be leveraged to further improve the performance of adversarial detection in secure machine learning. For example, methods from the area of copy-move forgery detection could be adapted to identify adversarial examples that modify large connected parts of an image and thus cannot be reliably detected by our method.

The bigger picture suggests not only that machine learning is useful in steganalysis and multimedia forensics, but also that secure machine learning should learn from these fields.

\bibliographystyle{IEEEtran}

\bibliography{literatur}

\end{document}